\documentclass[numberedappendix,appendixfloats,twocolappendix]{emulateapj}
\usepackage{natbib}
\usepackage{bm}
\usepackage{color}

\def\lya{Ly$\alpha$ }
\def\Msun{M_\odot}
\def\revision[#1]{{#1}}

\begin{document}

\title{On the decreasing fraction of Strong Ly$\alpha$ Emitters around \lowercase{$z$} $\sim$ $6$-$7$}

\author{Raphael Sadoun\altaffilmark{1}}
\author{Zheng Zheng\altaffilmark{1}}
\author{Jordi Miralda-Escud\'{e}\altaffilmark{2,3}}

\altaffiltext{1}{Department of Physics and Astronomy, University of Utah, 
115 South 1400 East, Salt Lake City, UT 84112, USA}
\altaffiltext{2}{Instituci\'{o} Catalana de Recerca i Estudis Avan\c{c}ats, 
Barcelona, Catalonia}
\altaffiltext{3}{Institut de Ci\`{e}ncies del Cosmos, Universitat de Barcelona (IEEC-UB), Barcelona, Catalonia}

\email{raphael.sadoun@utah.edu}

\begin{abstract}
 The fraction of galaxies with strong \lya emission has been observed to
decrease rapidly with redshift at $z \ga 6$, after a gradual increase
at $z< 6$. This has been interpreted as a hint of the reionization of the
intergalactic medium (IGM): the emitted \lya photons would be scattered by
an increasingly neutral IGM at $z>6$. We study this effect by modeling
the ionization and \lya radiative transfer in the infall region and the
IGM around a \lya emitting galaxy (LAE), for a spherical halo model with
the mean density and radial velocity profiles in the standard $\Lambda$CDM
cosmological scenario. We find that the expected fast increase of the
ionizing background intensity toward the end of the reionization epoch implies
a rapid evolution of halo infall regions from being self-shielded against
the external ionizing background to being mostly ionized. Whereas
self-shielded infall regions can scatter the \lya photons over a much
larger area than the commonly used apertures for observing LAEs, the same
infalling gas is no longer optically thick to the \lya emission line after
it is ionized by the external background, making the \lya emission more
compact and brighter within the observed apertures. Based on this simple
model, we show that the observed drop in the abundance of LAEs at $z>6$
does not imply a rapid increase with redshift of the fraction of the whole
IGM volume that is atomic, but is accounted for by a rapid increase of the
neutral fraction in the infall regions around galaxy host halos.
\end{abstract}


\section{Introduction}

Cosmic reionization corresponds to the last major phase transition of the universe during 
which the intergalactic medium (IGM) transitioned from a highly neutral to a highly ionized 
state. The detailed history of how reionization proceeded is still poorly 
constrained owing to the limited number of observational 
tools to probe neutral gas in the early universe \citep[e.g., reviews by][]{Miralda-Escude03,Barkana.Loeb07,Ferrara.Pandolfi14}. 
The Gunn-Peterson troughs \citep{Gunn65} observed blueward of the Ly$\alpha$ transition in the spectra of high-redshift quasars indicates that 
reionization was largely completed by $z\sim 6$ \citep[e.g.,][]{Fan.etal02,White.etal03,Fan.etal06,Becker.etal07,Becker.etal15}. 
Constraints on reionization at $z \ga 6$ can be obtained from measurements of the Thomson scattering optical depth, 
$\tau_{\rm ts}$, using cosmic microwave background (CMB) data. 
As $\tau_{\rm ts}$ depends on the number density of free electrons 
integrated along the line-of-sight, it can be used to infer a characteristic 
reionization redshift $z_{\rm reion}$ but is insensitive to the precise 
reionization history. The latest results from the polarization data of
the Planck satellite's Low Frequency Instrument and CMB lensing 
indicate a value $\tau_{\rm ts} = 0.066 \pm 0.013$ \citep{Planck15}, 
and those from the High Frequency Instrument give
$\tau_{\rm ts} = 0.055 \pm 0.009$ \citep{Planck16a},
corresponding to a mean reionization redshift of 
$z_{\rm reion} \sim$ 7.8--8.8 \citep{Planck16b}.

Ly$\alpha$ emitting galaxies, or Ly$\alpha$ emitters\footnote{\lya emitters usually only refer to
  galaxies \emph{selected} to have strong \lya emission in narrow-band surveys.
  In this paper, we use the term \lya emitters more generally for any galaxy with
  detectable \lya emission} (LAEs)
constitute a promising alternative for studying neutral gas at early
times and can provide important constraints on the late stages of
reionization at $z \sim 6-7$
\citep[see e.g., a recent review by][]{Dijkstra14}. LAEs are young
star-forming galaxies in which most of the ionizing photons emitted from
hot stars are converted to Ly$\alpha$ photons after recombinations in
the interstellar medium (ISM), resulting in strong Ly$\alpha$ emission. 
As such, they have been predicted to be primary targets in the search
for high-redshift galaxies \citep{Partridge.Peebles67}. After Ly$\alpha$
photons escape the interstellar medium around the young stars where they
are produced, they experience resonant scattering by neutral hydrogen
atoms in the surrounding IGM. The Ly$\alpha$ line emitted from these
galaxies therefore contains information on the state of the neutral gas
in their vicinity.
 
  Before reionization was complete, an absorption imprint
should be left on the Ly$\alpha$ emission line of LAEs because of the
remaining atomic hydrogen in the IGM. The damped absorption wings of IGM
regions with a high neutral fraction are expected to substantially
suppress the Ly$\alpha$ emission lines of galaxies behind them
\citep[e.g.,][]{Miralda.Rees98,Miralda-Escude98}.
Recent observations of Ly$\alpha$ emitting galaxies at high
redshift have indeed revealed a reduction in the visibility of LAEs
between $z\sim 6$ and $7$. Using ultra-deep 
narrow-band imaging with the Subaru telescope, \citet{Konno.etal14} found a rapid decline in the Ly$\alpha$ luminosity 
function (LF) of LAEs from $z=6.6$ to $7.3$.
Combined with evidence for no evolution in the ultra-violet (UV) continuum 
LF over the same redshift interval, they concluded that reionization is
likely not complete at $z\sim 7$ and that this may explain the sudden
decline of the \lya\ LF of LAEs.
The Ly$\alpha$ fraction $X_{\rm Ly\alpha}$, defined as
the fraction of objects with strong Ly$\alpha$ emission 
among Lyman Break Galaxies (LBGs), is slowly rising 
from $z\sim 3$ to $6$ \citep[e.g.,][]{Stark.etal11}, but then
decreases suddenly between $z\sim 6$ and $7$ 
\citep{Fontana.etal10,Stark.etal10,Pentericci.etal11,Ono.etal12,Schenker.etal12,Caruana.etal14,Schmidt.etal16}. 
Although one can imagine evolution models for intrinsic galaxy
properties such as the escape fraction of ionizing photons or the dust
content that might explain this drop in observed Ly$\alpha$ emission
\citep[e.g.,][]{Dayal.Ferrara12}, the fraction $X_{\rm Ly\alpha}$ should
not decline in a synchronized way for all galaxies in the Universe over
a narrow time interval (note that $\Delta z = 1$ corresponds to less
than $\sim 200\,{\rm Myr}$ at $z\sim 6$, or $\sim$ 20\% of the age of the
Universe at that epoch), so this decline has naturally been interpreted
as a signature of the increase in the neutral gas fraction in the IGM
towards high redshift. 

  The main difficulty with the simple scenario where a smooth IGM at the
end of the reionization epoch is causing this drop of LAEs is that,
adopting a simple attenuation model for the transmission of the 
Ly$\alpha$ line through the intervening IGM, the observed drop in
$X_{\rm Ly\alpha}$ between $z=6$ and $z=7$ implies a rapid evolution
in the volume-averaged neutral fraction $\left<x_{\rm HI}\right>$ of
several tens of percent 
\citep{Pentericci.etal11,Dijkstra14}. This would demand a late
and very sudden reionization scenario, implying
a surprisingly rapid rise in the emission rate of ionizing photons, and
in tension with the Thomson scattering optical depth measurements 
by the {\it Wilkinson Microwave Anisotropy Probe} (WMAP; $z_{\rm reion} \sim 10.5$ derived in \citealt{Hinshaw.etal13}), although
the latest results from Planck \citep{Planck15,Planck16a,Planck16b} have 
eased the tension
with a late reionization.  

  For this reason, alternative scenarios have been proposed to explain
the reduction in Ly$\alpha$ flux at $z \sim 6-7$ without requiring large
variations in the neutral gas content of the IGM over this redshift
interval.

\citet{Bolton.Haehnelt13} suggested that the transmission of the
Ly$\alpha$ line might be significantly reduced due to the presence of
relatively dense and neutral gas absorbers that are 
self-shielded against ionizing radiation. The size of
these absorbers is reduced as the intensity of the ionizing UV radiation
background rises. During the late stages of reionization, the 
mean free path of ionizing UV photons can increase rapidly as the
self-shielded absorbers shrink, causing a rapid change in the ionizing
background intensity \citep{Miralda-Escude.etal00,Giallongo15,Madau15,Mitra16,Munoz.etal16}. 
As a consequence, the Ly$\alpha$ transmission through the intervening
IGM might be reduced significantly without requiring a large change in
$\left<x_{\rm HI}\right>$ from $z=6$ to $z=7$. However, using results
from reionization simulations, \citet{Mesinger.etal15} showed that these
self-shielded absorbers cannot fully account for the total IGM opacity
required to explain the observed drop in $X_{\rm Ly\alpha}$. 

  In general, the resonant nature of the Ly$\alpha$ line makes it
difficult to infer the neutral state of the IGM surrounding LAEs
directly from the evolution of $X_{\rm Ly\alpha}$, as radiative transfer
effects can significantly alter the transmission of the line
\citep[e.g.,][]{Zheng.etal10}.
In this paper, we aim to further explore the impact of a rapidly 
evolving ionizing UV background intensity on the visibility of LAEs at
$z\sim 6-7$, taking into account radiative transfer effects. 
Compared to the previous scenarios mentioned above, 
we focus our investigation on how the distribution of neutral gas surrounding the LAEs \emph{themselves} 
is affected by the local UV background when taking into account self-shielding effects. 
Furthermore, we calculate the full radiative transfer of Ly$\alpha$ photons using a Monte-Carlo approach 
in order to accurately predict the Ly$\alpha$ properties resulting from the transmission 
through this gas. For this purpose, we use an analytical description to model the gas distribution 
and kinematics around LAEs at high-redshift. As the radiative transfer of the Ly$\alpha$ line can be 
significantly modified depending on the gas kinematics, we also explicitly consider inflow of gas onto 
the host LAE halo. 

The paper is organized as follows. In Section \ref{sec:model}, we 
describe our model for the gas distribution around high-$z$ LAEs, and present the 
details of the self-shielding and Ly$\alpha$ radiative transfer calculations. 
The Ly$\alpha$ properties of our modeled LAEs as well as the main results on 
the evolution of the Ly$\alpha$ fraction are presented 
in Section \ref{sec:results}. Finally, we discuss the 
implications of our results and conclude in Section 
\ref{sec:conclusion}.

\section{LAE model} 
\label{sec:model}  

In this section, we start by presenting the analytical model we use 
for the density and velocity of the gas surrounding LAEs. 
Then, we describe how the self-shielding effect is taken into 
account to calculate the corresponding neutral gas distribution in the presence 
of an external ionizing UV background. Finally, we present the radiative transfer calculations 
used to predict the Ly$\alpha$ properties from our model.

\subsection{Gas Density and Velocity}

  Our model assumes that an LAE is in the center of a dark matter halo
of virial mass $M_h$, with a gas distribution that is modeled in terms
of the dark matter distribution. We assume a dark matter density 
profile within the virial radius of the halo, $r_h$,
described by the Navarro-Frenk-White (NFW) profile \citep{Navarro97},
\begin{equation}
\rho_{\rm \textsc{\tiny NFW}} (r) = \frac{\delta_c \rho_{\rm crit}(z)}{r/r_{s}(1+r/r_{s})^2},
\label{eq:rho_NFW}
\end{equation}
where $\delta_{c}$ is a characteristic overdensity, $r_{s}$ is the scale 
radius of the halo, $\rho_{\rm crit}(z)=3H(z)^2/8\pi G$ is the critical
density of the universe at redshift $z$, and $H(z)$ is the Hubble
constant at redshift $z$. Throughout this paper, we adopt the WMAP-9
cosmology \citep{Hinshaw.etal13} with $\Omega_m = 0.28$,
$\Omega_\Lambda = 0.72$, $\Omega_b = 0.046$ and $H_0= 70\, {\rm km/s/Mpc}$.

Both parameters $\delta_{c}$ and $r_{s}$ can be expressed in 
terms of the halo virial mass $M_{h}$ and concentration parameter 
$c_{\rm \textsc{\tiny NFW}} \equiv r_{h}/r_{s}$, where $r_{h}$ is the halo 
virial radius. In this paper, halos are defined as to have a mean density
of $\Delta \times \rho_{\rm crit}(z)$ within the virial radius $r_h$, which
gives
\begin{equation}
r_{h} = \left(\frac{3M_{h}}{4\pi \Delta \rho_{\rm crit}(z)}\right)^{1/3}
\label{eq:Rvir}
\end{equation}
and
\begin{equation}
\delta_c = \frac{\Delta}{3}\frac{c_{\rm \textsc{\tiny NFW}}^3}
{\ln(1+c_{\rm \textsc{\tiny NFW}})-c_{\rm \textsc{\tiny NFW}}/(1+c_{\rm \textsc{\tiny NFW}})}.
\label{eq:delta_c}
\end{equation}
We choose $\Delta = 18\pi^2$ for the density contrast 
as a good approximation to the value predicted by 
the spherical collapse model for a flat $\Lambda$CDM cosmology 
at $z \gtrsim 6$ \citep{Bryan.Norman98}. 
We fix $c_{\rm \textsc{\tiny NFW}} = 2$ (about that 
of a Milky-Way sized halo at $z \sim 6-7$) and neglect its weak dependence on 
halo mass. %As we shall see now, the gas density profile is modified from the
%NFW one and the Ly$\alpha$ properties of our model LAEs depend 
%mostly on the gas distribution on scales larger than the halo virial radius,
%so small variations in the halo density profile do not significantly affect 
%our results.

  The gas distribution inside the halo should differ from that of the dark 
matter for several reasons. First, for constant gas temperature, the gas
pressure flattens the distribution near the center, compared to the
diverging NFW profile. \revision[Second, the gas can cool, settle in a disk and form stars,
which will affect the rest of the gas due to ionization and supernovae. Moreover,
the neutral gas distribution, which is the one relevant for \lya transfer,
is different from the total gas one because of photoionization and density
and temperature variations. These effects are highly complex but, as we shall
see below, the \lya properties of our model LAEs depend mostly on the gas
distribution outside the halo virial radius. We therefore ignore these
complications at small radius, and simply impose a constant density core of
radius $r_{\rm core}=\alpha r_h$. Exterior to this core radius, we assume
the gas follows the NFW profile with density $\rho_{\rm gas}= f_b \rho_{\rm NFW}(r)$, where
$f_b=\Omega_b/\Omega_m$ is the global baryon fraction, and interior to the
core radius we have $\rho_{\rm gas}= f_b \rho_{\rm NFW}(r_{\rm core})$.
In our fiducial model, we use $\alpha = 0.25$ as a rough estimate for the scale 
at which the gas stops following the underlying dark matter distribution.
We examined the effect of varying the value of $\alpha$ (Appendix \ref{sec:appendix}) and found 
that it has little impact on our results since, again, the Ly$\alpha$ properties are mainly affected by 
the gas distribution outside of the halo that suffers from the self-shielding correction.]

The gas distribution outside the halo and in the IGM is assumed to trace that 
of the dark matter. To describe the dark matter distribution, we apply 
the infall model of \citet{Barkana04}, which predicts the average density and 
velocity profile around virialized structures of mass $M_h$ at a given 
redshift $z$ in the presence of infalling matter. Briefly, the model 
calculates the initial density profile around overdense regions in the 
density field linearly extrapolated to the present day using the extended 
Press-Schechter formalism \citep{Press74,Bond91}. 
The relation to the density and velocity profiles around virialized halos at a 
given redshift $z$ is then established based on the spherical collapse, 
taking into account the effect of matter infall onto the halo. 
The model also predicts the $1\sigma$ scatter around the average 
profiles expected in the matter distribution around halos, which allows us 
to quantify the variations in the Ly$\alpha$ properties caused by the 
distribution of gas environment in the IGM.
Note that this model has also been 
used by \citet{Dijkstra.etal07} to model the IGM to study the \lya 
transmission at $z \ga 6$.

At a certain radius $r_{\rm eq}$, the matter density $\rho_{\rm infall}$ given 
by the infall model reaches that given by the NFW profile 
$\rho_{\rm \textsc{\tiny NFW}}$. 
We find that $r_{\rm eq}\sim 2 r_h$ for halo masses 
in the range $10^{10}$--$10^{12}\Msun$ considered in this work. The gas 
density profile in our model is then set to be $\rho_{\rm gas}=f_b 
\rho_{\rm \textsc{\tiny NFW}}(r_{\rm core})$ for $r\leq r_{\rm core}$, 
$f_b \rho_{\rm \textsc{\tiny NFW}}(r)$ for $r_{\rm core} < r \leq r_{\rm eq}$, 
and $f_b \rho_{\rm infall}(r)$ for $r>r_{\rm eq}$. We also set the
maximum radius of the gas halo to be $10r_{h}$, for the purpose of
computing the emerging \lya spectrum.

For the velocity distribution of gas, we assume that it follows that of the 
dark matter outside of the virial radius of the halo, which accounts for the 
gas infall. Inside the virial radius, we assume that the gas is supported by 
dispersion with no bulk motion. We impose a smooth transition between the 
dispersion-dominated region (inside the virial radius) and the infall region 
(the surrounding IGM) and express the peculiar velocity $v_p$ of the gas as
\begin{equation}
v_p(x) = \frac{v_{\rm infall}(x)}{1+e^{-w(x-x_0)}},
\label{eq:vpec}
\end{equation}
where $x = r/r_{h}$. The parameters $w$ and $x_0$ control respectively the 
width 
and location of the transition between the two regimes. In the present model, 
we use $w = 20$ and $x_0 = 1$, which produces a sharp transition at the virial 
radius. For the total velocity $v_{\rm tot}$ of the gas, the 
contribution from the Hubble flow $v_{\rm Hubble}$ is added, i.e., 
$v_{\rm tot} = v_p + v_{\rm Hubble}$.

\begin{figure}
\plotone{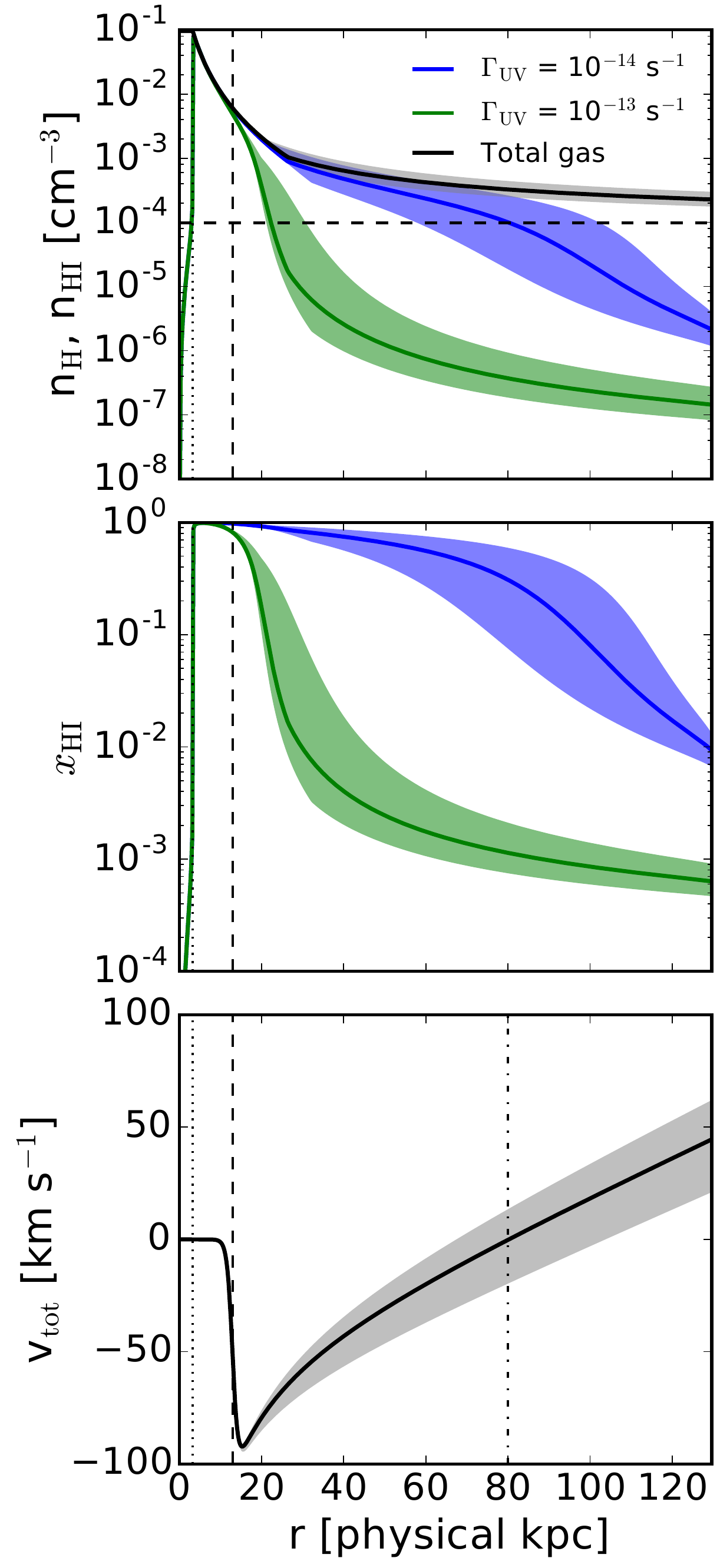}
\caption{
Gas density and velocity profiles of our fiducial LAE model for a halo of 
mass $M_{h}=10^{10.5}\,\mathrm{M}_\sun$ and virial radius $r_{h}\sim 13\, \mathrm{kpc}$ at $z=7$. 
Top panel shows the density profile of the total (black line) and neutral 
(blue and green lines) hydrogen gas. The neutral gas profiles are obtained
by calculating the self-shielding effects in the presence of a uniform 
external ionizing background, with corresponding photoionization rates 
$\Gamma_{\rm UV}=10^{-13}\,\mathrm{s}^{-1}$ (green) and $\Gamma_{\rm UV}=10^{-14}\,\mathrm{s}^{-1}$ (blue), respectively. 
The horizontal dashed line represents the mean baryonic density of the universe. 
The middle panel shows the corresponding neutral hydrogen fractions $x_{\rm HI} \equiv n_{\rm HI}/n_{\rm H}$. 
The bottom panel shows the total bulk velocity of the gas, where the
vertical dot-dashed line marks the turnaround radius (see text). 
In all panels, the shaded region around each average profile shows the scatter
resulting from the distribution of the IGM environment predicted 
by the infall model. The vertical dotted and dashed lines represent 
respectively the core radius of the gas distribution and the virial radius 
of the halo. 
}
\label{fig:gas_profiles}
\end{figure}

The top and bottom panels of Figure~\ref{fig:gas_profiles} show the resulting 
density and velocity profiles for a $z=7$ halo of mass $M_h = 10^{10.5}\Msun$, 
the typical mass of LAE host halos at $z \ga 6$ as inferred from clustering analysis \citep{Ouchi.etal10,Sobacchi.Mesinger15}. 
The virial radius of the halo at this redshift is $r_{h} \sim 13\,\mathrm{kpc}$ (physical). 
The vertical dotted and dashed lines represent respectively the core and virial radius of the halo. 
The shaded regions represent the scatter expected in the total gas density and velocity profiles as predicted by the 
infall model. 
The total mean gas density is plotted in the top panel as the solid black 
curve, while the horizontal dashed line marks the mean baryonic density of 
the Universe.
Note that the density predicted by the infall model falls off 
asymptotically towards the mean value.
The total IGM radial velocity is negative (indicating infall) below the
turnaround radius, which occurs at $r\sim 6r_{h}$ (indicated by the vertical
dot-dashed line in the bottom panel; this agrees with results of numerical
simulations, e.g., in \citealt[]{Meiksin.etal14}),
and asymptotically approaches the Hubble flow at larger radius. 

To summarize, at a given redshift $z$, our model describing the gas 
density and velocity distribution in the halo and the IGM depends only on a 
single parameter, the halo virial mass $M_{h}$. Throughout the paper, we 
assume a fixed value of $z=7$ for all the calculations. The main reason for 
this choice is to isolate the effects of varying the intensity of the 
ionizing background while keeping the other parameters fixed. In practice,
through tests we have verified that including the redshift dependence in our 
model does not affect our main conclusions on the evolution of the fraction of strong LAEs. 
We convert the gas density to hydrogen number density by 
$n_{\rm H} = X_{\rm H}\rho_{\rm gas}/m_{\rm H}$, where $X_{\rm H} = 0.76$ is the hydrogen mass fraction and $m_{\rm H}$ is the mass of the hydrogen atom. 

\subsection{Self-Shielding Calculation}

For the \lya radiative transfer, we need to know the neutral hydrogen 
distribution. With the gas distribution in Section~2.1, we solve
the neutral hydrogen distribution by accounting for the self-shielding effects 
in the presence of a uniform external ionizing background \revision[as well as the ionizing flux 
from the LAE at the center], following the approach in \citet{Zheng.Miralda02a} with an iterative procedure \citep{Tajiri.Umemura98}. 
The neutral fraction $x_{\rm HI}(r) \equiv n_{\rm HI}(r)/n_{\rm H}(r)$ at a 
given distance $r$ from the center of the cloud is found by solving the 
photonionization equilibrium equation,
\begin{equation}
x_{\rm HI}(r)\Gamma_{\rm tot, ss}(r) = \alpha_{\rm B}(T)\left[1-x_{\rm HI}(r)\right] n_{e},
\label{eq:photoequi}
\end{equation}
where \revision[$\Gamma_{\rm tot, ss}$ is the \emph{total} attenuated photoionization rate], 
$\alpha_{\rm B}$ is the case-B recombination coefficient at temperature $T$, 
and $n_{e}$ is the electron number density. We assume a temperature of 
$T=10^4\, \mathrm{K}$ ($\alpha_{\rm B}  = 2.35\,\times\,10^{-13}\,\mathrm{cm}^{3}\,\mathrm{s}^{-1}$). For the electron density, we account for the contribution
from singly ionized helium with the same neutral fraction as hydrogen. With 
a helium mass fraction of $Y_{\rm He}=0.24$, we have 
$n_{e} = \left(1-x_{\rm HI}\right)0.82\rho_{\rm gas}/m_{\rm H}$. 

\revision[The total photoionization rate is the sum of the contributions from the central 
LAE source and the external UV background, $\Gamma_{\rm tot, ss} = \Gamma_{\rm LAE, ss}+ \Gamma_{\rm UV, ss}$. 
Assuming that ionizing photons are escaping isotropically from the LAE, the 
photoionization rate $\Gamma_{\rm LAE, ss}$ at a distance $r$ can be expressed as]

\begin{equation}
\Gamma_{\rm LAE, ss}(r) = \int_{\nu_{\rm L}}^{\infty}\frac{f_{\rm esc}L_{\nu}e^{-\tau_{\nu, \rm cen}}}{4\pi r^2 h\nu}a_{\nu}d\nu,
\label{eq:PIrate2}
\end{equation}

\revision[
where $\nu_{\rm L}$ is the \ion{H}{1} Lyman limit frequency, $\tau_{\nu, \rm cen}$ is the 
photoionization optical depth from the \emph{center of the gas cloud} to $r$, 
$a_\nu$ is the photoionization cross-section of hydrogen at frequency $\nu$, 
$L_{\nu}$ is the specific (ionizing) luminosity of the LAE and $f_{\rm esc}$ is the escape fraction of ionizing photons.
We assume that the galaxy spectrum is of the form $L_{\nu} \propto \nu^{-\beta}$ for $\nu > \nu_{\rm L}$ and that 
the total emission rate of ionizing photons $Q$ is related to the instantaneous star formation rate (SFR) and 
the metallicity $Z$ (in solar units) by]

\begin{equation}
\log_{\rm 10} Q = 53.8 + \log_{\rm 10}(\mathrm{SFR}/\Msun\,\mathrm{yr}^{-1}) - 0.0029(9+\log_{\rm 10}Z)^{2.5},
\label{eq:ion_lum}
\end{equation} 

\revision[which assumes constant SFR and a Salpeter IMF \citep{Schaerer2003,Dijkstra.etal07}. 
We relate SFR to halo mass using the relation \citep{Zheng.etal10}]
\begin{equation}
\mathrm{SFR} = 0.68[M_{h}/(10^{10} h^{-1} \mathrm{M}_\sun)]\, \mathrm{M}_\sun\, \mathrm{yr}^{-1}.
\label{eq:sfr_Mh}
\end{equation} 
\revision[For our fiducial model, we assume solar metallicity ($Z=1$), an escape
fraction of ionizing photons from the central star-forming region of
$f_{\rm esc}=0.1$, and $\beta=3$, which results in a total emission rate of
$Q f_{\rm esc} \simeq 2\times 10^{52}\, {\rm photons}/s$ for a halo of
mass $M_h = 10^{10.5}\,\Msun$, using equations \ref{eq:ion_lum} and \ref{eq:sfr_Mh}.]

Given the intensity $I_{\nu}$ of the external ionizing background, 
the photoionization rate \revision[$\Gamma_{\rm UV, \rm ss}$] is computed as
\begin{equation}
\Gamma_{\rm UV, ss} = \int_{4\pi}d\Omega\,\int_{\nu_{\rm L}}^{\infty}\frac{I_{\nu}e^{-\tau_{\nu, \rm out}}}{h\nu}a_{\nu}d\nu,
\label{eq:photorate}
\end{equation}
\revision[where $\tau_{\nu, \rm out}$ is the photoionization optical depth 
from \emph{outside the cloud}] to $r$ along a given direction.
The integration over the solid angle accounts for ionizing
photons coming from all directions to the position at $r$. 
We assume that the ionizing background intensity $I_\nu$ is a constant, 
denoted as $I_{\nu,0}$, between $\nu_{\rm L}$ and $4\nu_{\rm L}$, and zero 
above $4\nu_{\rm L}$ (the \ion{He}{2} Lyman limit). The constant $I_{\nu,0}$ 
is a parameter of our model and is chosen such that it gives the desired value 
of the photoionization rate $\Gamma_{\rm UV}$ of the ionizing background when computed 
in the optically thin regime ($\tau_{\nu}<<1$). 
Since the redshift evolution of $\Gamma_{\rm UV}$ is still poorly constrained at $z\ge 6$, 
we consider two typical cases in our fiducial model with $\Gamma_{\rm UV}=10^{-13}\,\mathrm{s}^{-1}$ and 
$10^{-14}\,\mathrm{s}^{-1}$, corresponding approximately to the expected 
evolution between $z=6$ and $z=7$ \citep{Fan.etal06,Becker13,Giallongo15,
Madau15,Mitra16,Munoz.etal16}.

The photoionization equilibrium equation is solved using an iterative method. 
First, the neutral fraction profile is initialized assuming the system to be 
optically thin to ionizing photons [$\tau_\nu << 1$ in 
equation~(\ref{eq:photorate})]. Then, at each consecutive iteration, the value 
of the optical depth $\tau_{\nu}$ is recalculated in all directions according 
to the updated neutral gas distribution and the neutral fraction is solved 
using equation~(\ref{eq:photoequi}). The process is repeated until the neutral 
fraction has converged within a fractional error of less than $10^{-5}$. 

The top and middle panels of Figure~\ref{fig:gas_profiles} show respectively 
the neutral gas density and the corresponding neutral fraction profiles 
obtained for the $10^{10.5}\,\Msun$ halo.
The green and blue curves are for $\Gamma_{\rm UV}=10^{-13}\,\mathrm{s}^{-1}$ 
and $\Gamma_{\rm UV}=10^{-14}\,\mathrm{s}^{-1}$, respectively. The shaded regions show 
the expected scatter around the average, coming directly from the variation 
in the IGM environment predicted by the infall model. Note that we do not 
consider variations in the gas distribution inside the halo since, as 
mentioned previously, these have little effect on the resulting Ly$\alpha$ 
properties. 

As expected, varying the ionizing background by one order of magnitude 
has a significant impact on the neutral gas distribution owing to the 
self-shielding effects. The main difference is seen in the ionization state 
of the infall region surrounding the halo for $r_{h} \la r \la 6r_{h}$. 
For $\Gamma_{\rm UV}=10^{-14}\,\mathrm{s}^{-1}$, the gas in this region remains 
highly neutral, while it is highly ionized ($x_{\rm HI} \sim 10^{-3}-10^{-2}$ 
at $r>2r_h$) for $\Gamma_{\rm UV}=10^{-13}\,\mathrm{s}^{-1}$. This can also be 
quantified in terms of the transition radius where $x_{\rm HI}=0.5$.
With the low UV background, this transition happens close to the turnaround
radius, at $r\sim 5r_{h}$, while with the high UV background, the
transition occurs in the vicinity of the halo virial radius, at
$r\sim 1.3r_{h}$.

\revision[In both cases, the transition radius is larger than the virial radius of
the halo, so the gas inside the virial radius is self-shielded against the
external ionizing backgroung. Therefore, the gas in the infall region
is the one that responds to a change in the external ionizing intensity
for this range of $\Gamma_{UV}$, and determines the transfer of the
emerging \lya radiation and the visibility of LAEs.]

\revision[ The presence of the central source of ionizing photons does not substantially
modify the neutral gas distribution of this infall region, for the ionizing
central luminosity we have assumed. The reason is that the total
recombination rate in the core region within $r_{\rm core}=0.25 r_h$ is
$(4\pi/3)\, r_{\rm core}^3 \alpha_B \, 0.82 n_H\simeq 2\times 10^{52}\,
{\rm photons}/s$, comparable to the total emission $Q f_{\rm esc}$. Therefore,
the central source simply creates a central HII region that barely suffices
to ionize the core, and does not affect the ionization structure of the
outer regions, as seen in Figure \ref{fig:gas_profiles}. Had we assumed a larger central
ionizing luminosity, the central source would of course have ionized
completely the surrounding gas and the \lya photons would then emerge with
little scattering, drastically changing our predictions.
Our model is therefore valid for halos in which the central
source is not very luminous. In practice, we expect that there is a minority
of halos in which a highly luminous source (either a quasar or a galaxy with
an extremely high star-formation rate) completely ionizes the gas surrounding
the halo (and then contributes to the reionization of the intergalactic
medium), while in the majority of halos the internal sources are not
capable of ionizing all the halo gas and the infall region needs to be ionized
by the external ionizing background. As long as the full ionization of the
gas inside the virial radius does not occur, the details of the gas
distribution assumed within $r_h$ should not impact our results on the
transmission of the \lya emission line (see Appendix \ref{sec:appendix}).]

\subsection{\lya Radiative Transfer}

\begin{figure*}
\epsscale{1.1}
\plottwo{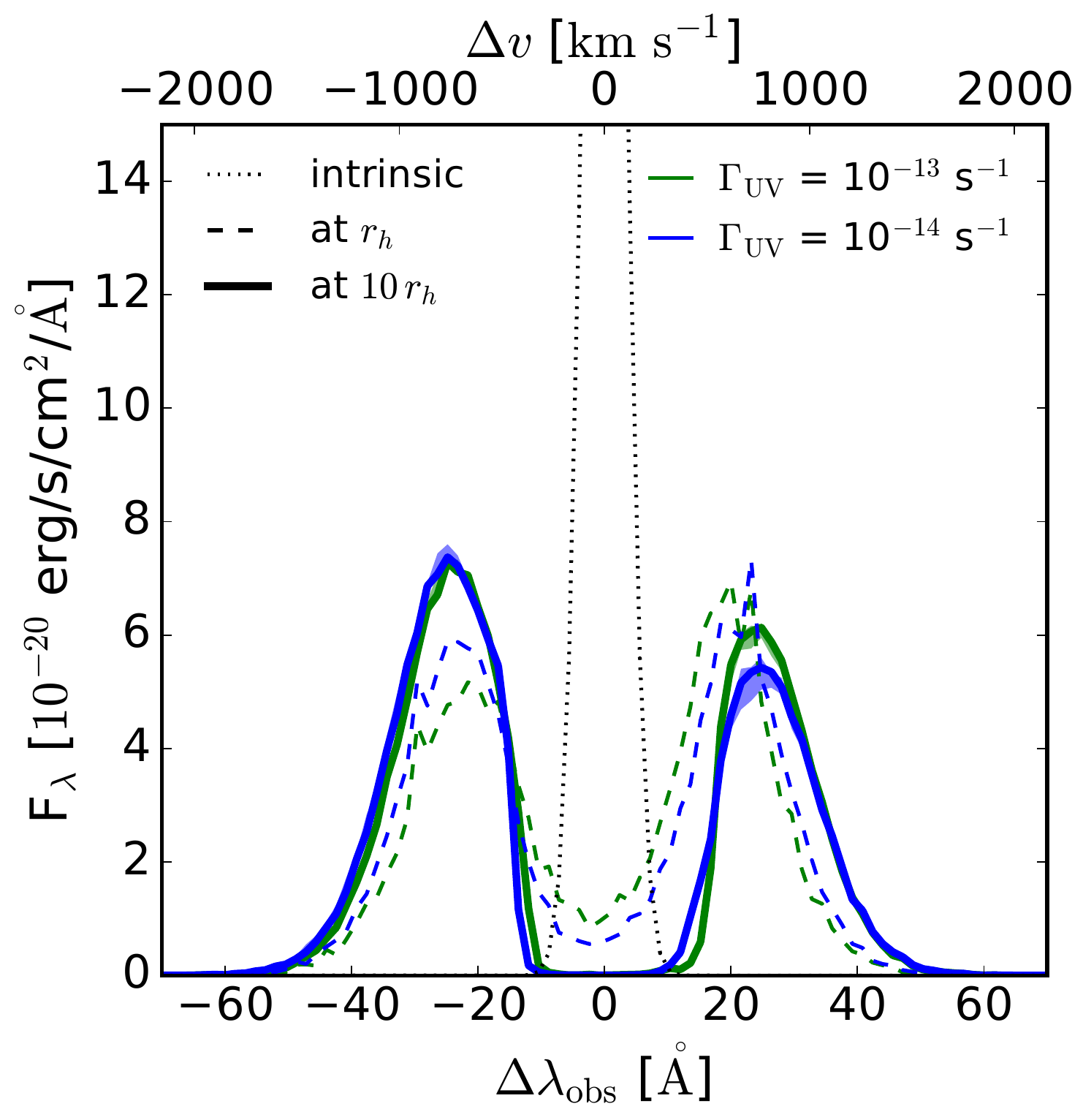}{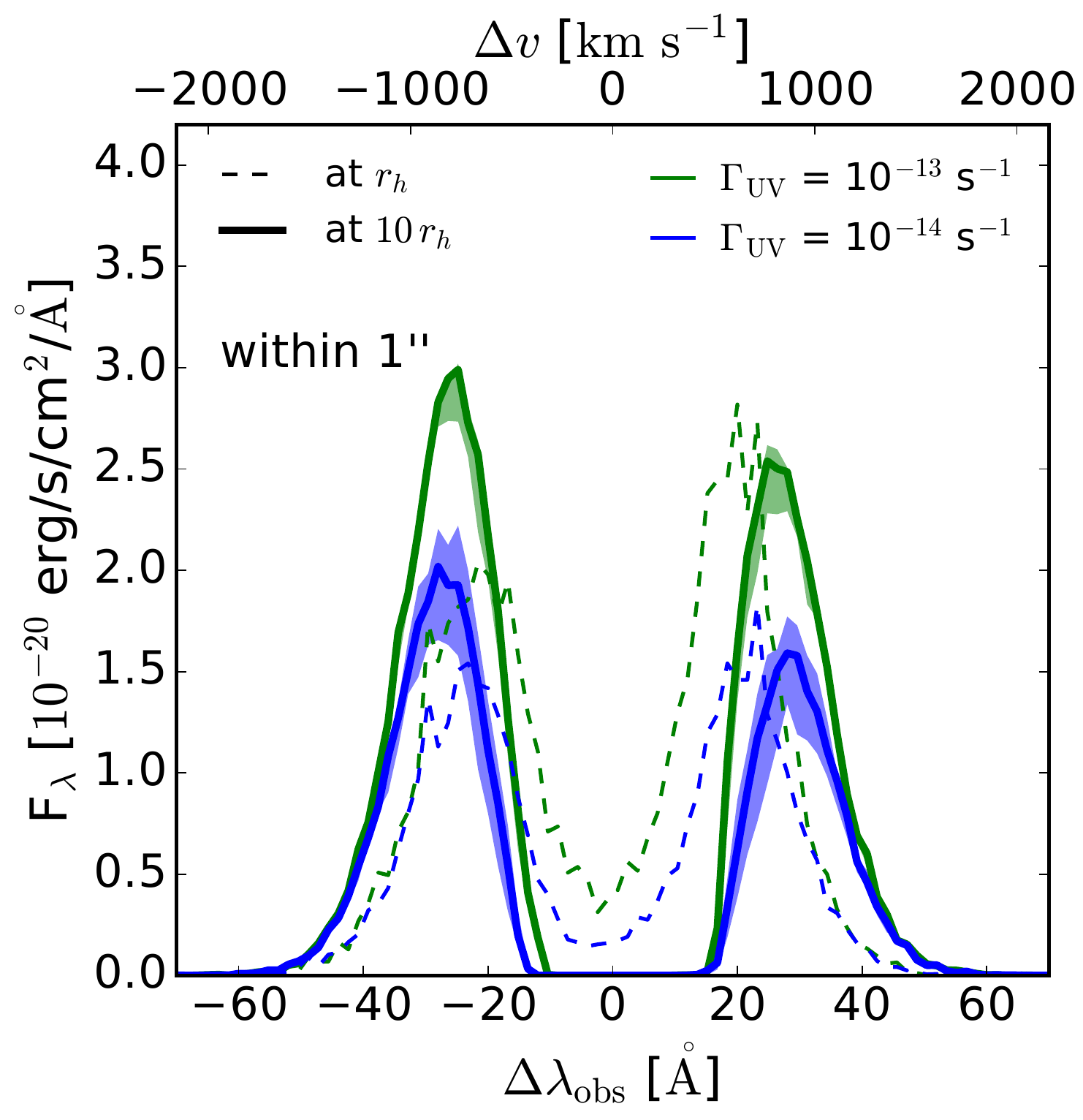}
\caption{
Flux per unit wavelength of the \lya emission of a model LAE in halos of
$M_h=10^{10.5}\Msun$, at $z=7$.
The spectra are obtained either for all photons (\emph{left}) or only for 
photons escaping within projected radius of $1\arcsec$ from the central source 
(\emph{right}). \revision[The bottom $x$-axis shows the offset with respect to the line 
center, expressed in the observer frame in wavelength units 
$\Delta{\lambda_{\rm obs}}=(1+z)\Delta{\lambda}$, while the top axis shows the offset 
in velocity units $\Delta v = c\,\Delta{\lambda} / 1215.67\AA$.] % with $\lambda_{\rm Ly\alpha} = 121.67 \AA$ being the \lya wavelength.]
The solid lines are the predicted spectra when photons reach the outer radius
of our radiative transfer calculation at $10 r_h$,
while the dashed lines are the spectra of photons reaching the
virial radius of the halo $r_h$.
The dotted black line is the intrinsic spectrum that would be observed if 
no radiative transfer effects were taken into account.
The blue side of the spectra will be highly supressed through attenuation by
the IGM far away from the source, and only the red side of the spectra is left
to be observed (see text).}
\label{fig:spectra}
\end{figure*}

With the density and velocity distribution of the neutral hydrogen gas in the
model, we perform a \lya radiative transfer calculation using a modified
parallel version of the Monte-Carlo code from \citet{Zheng.Miralda02b}. 
The code numerically follows the trajectories of Ly$\alpha$ photons, which 
diffuse spatially and in frequency as they scatter in the model cloud.
Ly$\alpha$ photons are launched at line center from the center of the gas 
halo and travel along a random direction until they are scattered. 
The location of the scattering event is obtained by drawing a random optical 
depth $\tau$ (following an exponential distribution) and computing the 
corresponding distance in the chosen direction. After each scattering, a new  
frequency and propagation direction of the photon are computed based on the 
velocity of the atom that scatters the photon. This scattering process is 
repeated until the photon escapes the gas cloud, at which point we record 
its frequency, its direction of propagation, and the location of the last 
scattering. We refer the reader to \citet{Zheng.Miralda02b} for more details.

Following \citet{Zheng.etal10}, we relate the Ly$\alpha$ luminosity of our 
model LAEs \revision[to their SFR,] 
\begin{equation}
L_{\mathrm{Ly}\alpha} = 10^{42} [\mathrm{SFR}/(\mathrm{M}_\sun \mathrm{yr}^{-1})]\, \mathrm{erg}\, \mathrm{s}^{-1},
\label{eq:lya_lum}
\end{equation} 
\revision[where the SFR is related to halo mass using Equation \ref{eq:sfr_Mh}.]
Our results on the influence of varying the ionizing background on the LAE fraction
are not sensitive to the particular $L_{\mathrm{Ly}\alpha}-M_{h}$ relation, 
as shown later. The effective temperature for gas inside halos for the
purpose of computing the \lya radiative transfer is set to be
$10^6\, {\rm K}$.
This is different from the gas temperature
$T=10^4\, {\rm K}$ assumed for the self-shielding calculation, and is
introduced to account for a realistic level of turbulent fluid motions
in the infall region. 
One might expect variations in the predicted Ly$\alpha$ properties as
we change the gas temperature for the radiative transfer calculations,
given the impact of Doppler broadening on \lya radiative transfer. 
By performing tests with the temperature set to $T=10^4\, {\rm K}$, we found
that the decrease in Ly$\alpha$ flux caused by the change in
the UV background remains essentially the same as in our fiducial model
and thus our final conclusion is not substantially affected. 

We model LAEs in halos of $\log (M_h/\Msun)= 10.0$ to $\log (M_h/\Msun)=12.0$ 
with a step size of 0.5 dex. For each halo mass, we consider 3 different IGM 
environments corresponding respectively to the average and $\pm1\sigma$ scatter  
in the gas density and velocity profiles predicted from the infall model in order 
to compute the scatter in the results. The number of photons we have numerically followed for each case is 
$N = 10^5$. We present the results from the radiative transfer calculation in the next section.

\section{Results}
\label{sec:results}

We now study the Ly$\alpha$ spectra and surface brightness profiles of
our model LAEs, based on our radiative transfer calculations. Then, we
present the main result of this paper on the evolution of the Ly$\alpha$
fraction at $z \ga 6$.

\subsection{Ly$\alpha$ Spectra}
\label{subsec:spectra}

  We record the location of last scattering as well as the frequency and 
direction of all \lya photons when they escape the gas cloud at the
outer radius of our radiative transfer calculation, which we have set at
$10 r_h$. This information enables us to compute the \lya spectrum of an LAE
at any projected radius. The left panel of Figure~\ref{fig:spectra}
shows the flux per unit wavelength of the photons as they come out of
the cloud at $10 r_h$, as the thick solid lines. The results are shown
for a halo of mass $M_h=10^{10.5}\Msun$ at $z=7$, and for the neutral gas
distribution of the two ionizing background intensities that were used
in Figure~\ref{fig:gas_profiles}: $\Gamma_{\rm UV}=10^{-13}\, {\rm s}^{-1}$
(green lines) and $\Gamma_{\rm UV}=10^{-14}\, {\rm s}^{-1}$ (blue lines).

  To gain physical insight into the effects of scattering by the
hydrogen in the infall region on the emission line shape, we also record
the frequency of photons as they first move out of the virial radius of
the halo, $r_h$. These spectra are shown in the left panel of
Figure~\ref{fig:spectra} as the dashed lines, for the same two cases as
the solid lines. The black, dotted line is the intrinsic emission
profile of the central source assumed in our model.
Note that the photon luminosity of the central source is the same in all
cases, and all the emitted photons eventually escape after multiple
scatterings, so the area under all five curves in the left panel is the
same.

  The resulting scattered spectra have the characteristic double-peak
profile expected from a central point source within a static gas
distribution \citep{Neufeld90,Zheng.Miralda02b,Dijkstra06}.
When photons are scattered in a region with a negative radial velocity
gradient (in our case, a narrow range around $r_h$; see bottom panel
of Fig. \ref{fig:gas_profiles}), the red peak \revision[($\Delta\lambda > 0$)] acquires a greater
intensity; this is seen clearly in the model of $\Gamma_{\rm UV}=10^{-13}\,{\rm s}^{-1}$
in the left panel of Figure~\ref{fig:spectra} (dashed
green line), where the faster reduction of the neutral density with
radius compared to the $\Gamma_{\rm UV}=10^{-14}\, {\rm s}^{-1}$ case enhances
the peak asymmetry that is produced. At radius $r_h$, an important
fraction of the photons are still left between the two peaks. As photons
diffuse further out into the region of positive radial velocity
gradient, the effect is reversed and photons that are scattering between
the two peaks are then more likely to end up in the blue peak \revision[($\Delta\lambda < 0$)]. Overall,
the radiative transfer through the infall region of the halo shifts most
of the photons remaining near the central region of the line to the blue
peak, and pushes the red peak further to the red, suppressing its
intensity with respect to the blue one.

  The spectra shown in this left panel of Figure~\ref{fig:spectra} are
not what is directly observed. Two effects need to be taken into account
that further modify the spectrum. First, the apertures that are most
often used to measure the spectrum are much smaller than the angular
size of the outer radius of $10 r_h$ at which we compute the emerging
spectrum, which is $\sim 24\arcsec$ at $z=7$. Second, these emerging
photons will still undergo further scattering beyond the limiting
radius $10 r_h$ used in our calculation. In reality, photons in the
blue peak should not be observed because, as they travel to the
observer, they will shift to the \lya line center at a radius much
larger than $10 r_h$, and they will be scattered again and reemitted
from a region of much larger angular size, and therefore with a very
low surface brightness. On the other hand, photons in the red peak
should not be further scattered after they have moved beyond $10 r_h$.
We have performed tests by extending the size of the cloud of gas to
$30r_h$, which show that the blue component becomes highly 
suppressed, while the red component remains unchanged. Therefore we can
include this second effect by simply assuming that only the red peak is
observed, and the blue peak is completely suppressed by scattering in
the IGM that is in Hubble expansion around the halo at $r> 10 r_h$. 
\revision[Note that this simple treatment implicitly assumes that photons propagate 
in a smooth IGM after they reach the limit of the gas cloud in our model and does not take 
into account additional absorption from self-shielded (Lyman-limit) systems 
in a clumpy IGM at $r > 10 r_h$. For this reason, the apparent drop in \lya flux that we predict 
in our model (Section~\ref{subsec:lya_sb}) should be considered as a lower limit to the total 
\lya flux decrement produced by self-shielding effects.]

  To include the first effect, we now assume that the \lya emission is
observed with a circular aperture of radius $1\arcsec$, a typical
aperture used for source extraction in narrow-band high-$z$ LAE
surveys. For each photon that reaches the outer radius $10 r_h$, we
compute the projected radius after its last scattering,
$R=\sqrt{r_{\rm ls}^2 - (\bm{k} \cdot \bm{r}_{\rm ls})^2 }$, where 
$\bm{r}_{\rm ls}$ is the vector position (with respect to the halo
center) of the last scattering, and $\bm{k}$ is the unit vector along
the photon escaping direction. The spectra of photons with $R$ within
the $1\arcsec$ aperture are shown in the right panel of
Figure~\ref{fig:spectra}, for the two cases of $\Gamma_{\rm UV}$. Solid lines
are the spectra of these photons as they exit the outer radius
$10 r_h$, and dashed lines show the spectrum these photons had when
first moving out of $r_h$. Note that for this latter case, the
projected radius is still computed after the escape from $10 r_h$, and
only the frequency is recorded at $r_h$, simply to show how the
frequency of these photons has changed as a result of the radiative
transfer through the infall region. The area under the two green
curves is therefore the same, and so is the area under the two blue
curves, but these areas are lower than in the left panels by a factor
equal to the fraction of photons that are eventually emitted within
the aperture of radius $1\arcsec$.

  We can now clearly see the effect of having a self-shielded infall
region around the halo of a LAE. For the more intense ionizing
background (green lines), when self-shielding starts only near the
virial radius $r_h$, the photons that escape in the red peak are
unlikely to be scattered at a radius substantially larger than $r_h$
and as a result the flux within the $1\arcsec$ aperture is suppressed by
only a factor \revision[$\sim 2.5$] relative to the total flux in the red peak.
However, when the ionizing background is weaker and the entire infall
region is self-shielded (blue lines), most photons are scattered at a
larger projected radius and the red peak is suppressed by a factor
\revision[$\sim 4$]. We note that, in both cases, more than half of the photons
are scattered into the blue peak, similarly to the effect seen for all
the photons in the left panel, but the variation in this blue peak
fraction is small and is not related to the change in the \lya
visibility. The reason for the greater reduction of the observed \lya
line (i.e., the red peak) for the $\Gamma_{\rm UV} = 10^{-14}\, {\rm s}^{-1}$
case is the much more extended neutral hydrogen profile caused by
self-shielding, which spreads the \lya photons over a much larger area,
therefore reducing the surface brightness and the detected flux within
the small apertures used for observing the spectra. 

\revision[Our model predicts an apparent \lya line profile (the red peak of the spectra) with a velocity offset 
of $\sim 800-900\,\mathrm{km}\,\mathrm{s}^{-1}$ with respect to the systemic redshift of 
the LAE. This is larger than the observed \lya line profiles for which the offset is typically 
$\sim 100-450\,\mathrm{km}\,\mathrm{s}^{-1}$ at $z\sim 2$ \citep{Shibuya.etal14,Stark.etal14} and 
$\sim 200-500\,\mathrm{km}\,\mathrm{s}^{-1}$ at $z\ga 6$ \citep{Willott.etal15}. Even the velocity offsets 
in the predicted spectra emerging at $r_h$ are already larger compared to the observed values which indicates 
that the neutral gas density inside the halo is probably overestimated in our model. This could be due to the 
relatively low escape fraction of ionizing photons that we assume in our fiducial model ($f_{\rm esc}=0.1$). 
We have examined the effect of increasing the escape fraction by a factor of two in our model 
(see Appendix \ref{sec:appendix}) and found that it reduces the velocity offset to 
$\sim 550\,\mathrm{km}\,\mathrm{s}^{-1}$, bringing it much closer to the observed value but 
without affecting the apparent flux reduction in the red peak when varying $\Gamma_{\rm UV}$.]

\subsection{\lya Surface Brigthness Profiles}
\label{subsec:lya_sb}

  We now examine in more detail the variation in the \lya surface
brightness profile as the ionizing background decreases, producing a
self-shielded infall region.

\begin{figure}
\epsscale{1.2}
\plotone{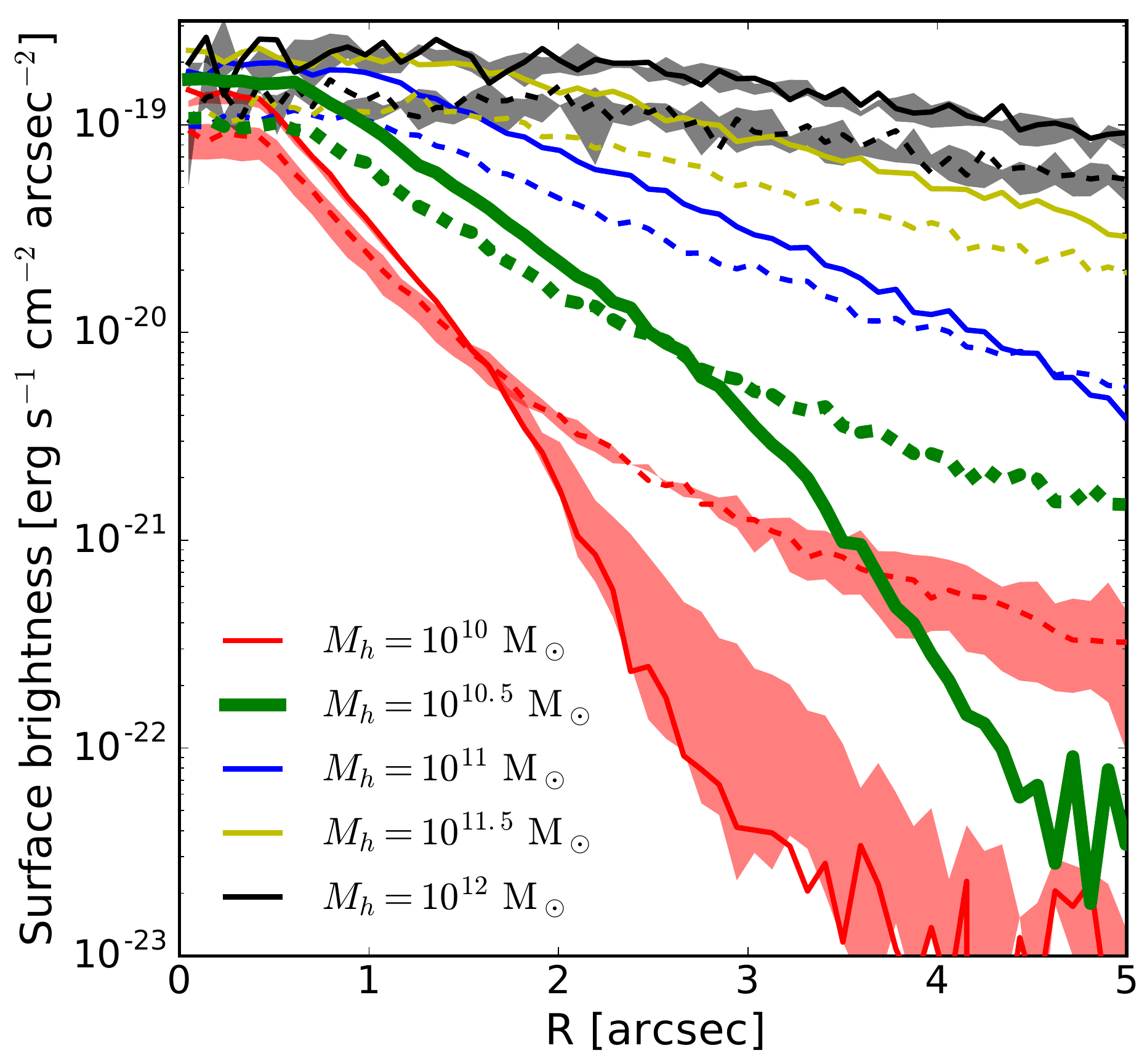}
\caption{
\lya surface brightness profiles for model LAEs residing in halos of 
various masses.
Solid lines and dashed lines are the profiles obtained for 
$\Gamma_{\rm UV}=10^{-13}\,\mathrm{s}^{-1}$ and $\Gamma_{\rm UV}=10^{-14}\,\mathrm{s}^{-1}$,
respectively. For clarity, the scatter (shaded regions) resulting from 
variable IGM environments (see Fig.~\ref{fig:gas_profiles})
is shown only for the lowest and highest halo mass.
}
\label{fig:sb_profiles}
\end{figure}

  Figure~\ref{fig:sb_profiles} plots the \lya surface brightness
profiles for different halo masses spanning the range
$10^{10}-10^{12}\Msun$. The cases for $\Gamma_{\rm UV}=10^{-13}\,\mathrm{s}^{-1}$ and $\Gamma_{\rm UV}=10^{-14}\,
\mathrm{s}^{-1}$ are represented as solid and dashed lines, 
respectively. The curves show the average profiles, and the expected
scatter around them caused by variations of the IGM gas distribution is
shown only for the lowest and highest halo mass to avoid excessive
cluttering.

  As the ionizing background increases from $\Gamma_{\rm UV}=10^{-14}\,
{\rm s}^{-1}$ to $10^{-13}\, {\rm s}^{-1}$, the gas is ionized at
smaller radius (see Figure~\ref{fig:gas_profiles}) and \lya photons
suffer less spatial diffusion before escaping the cloud. This causes a
steeper surface brightness profile, and a higher observed flux within
a small aperture near the center.

The surface brightness profile shows a clear dependence on halo mass. As halo 
mass increases, the self-shielding radius shifts toward large radius and the 
neutral hydrogen column density within the self-shielded region increases. As 
a consequence, \lya photons experience more scatterings and diffuse more 
spatially before entering the highly ionized, low density region. For 
$M_h=10^{12}\Msun$, the surface brightness profile becomes very flat out to
more than 5 arcseconds, which is not consistent with observed LAEs. However,
most LAEs at $z>6$ are probably not hosted in these extremely massive
halos. Our model is also unrealistic for the region inside the virial
radius, where turbulent motions and gas clumpiness are likely to be present
which would affect radiative transfer and the surface brightness profile. 
Nevertheless, the surface brightness profiles for the two cases of
$\Gamma_{\rm UV}$ values in Figure~\ref{fig:sb_profiles}
provides a reasonable indication for the ratio of \lya emission intensities
to be expected in the two models.

\begin{figure}
\epsscale{1.2}
\plotone{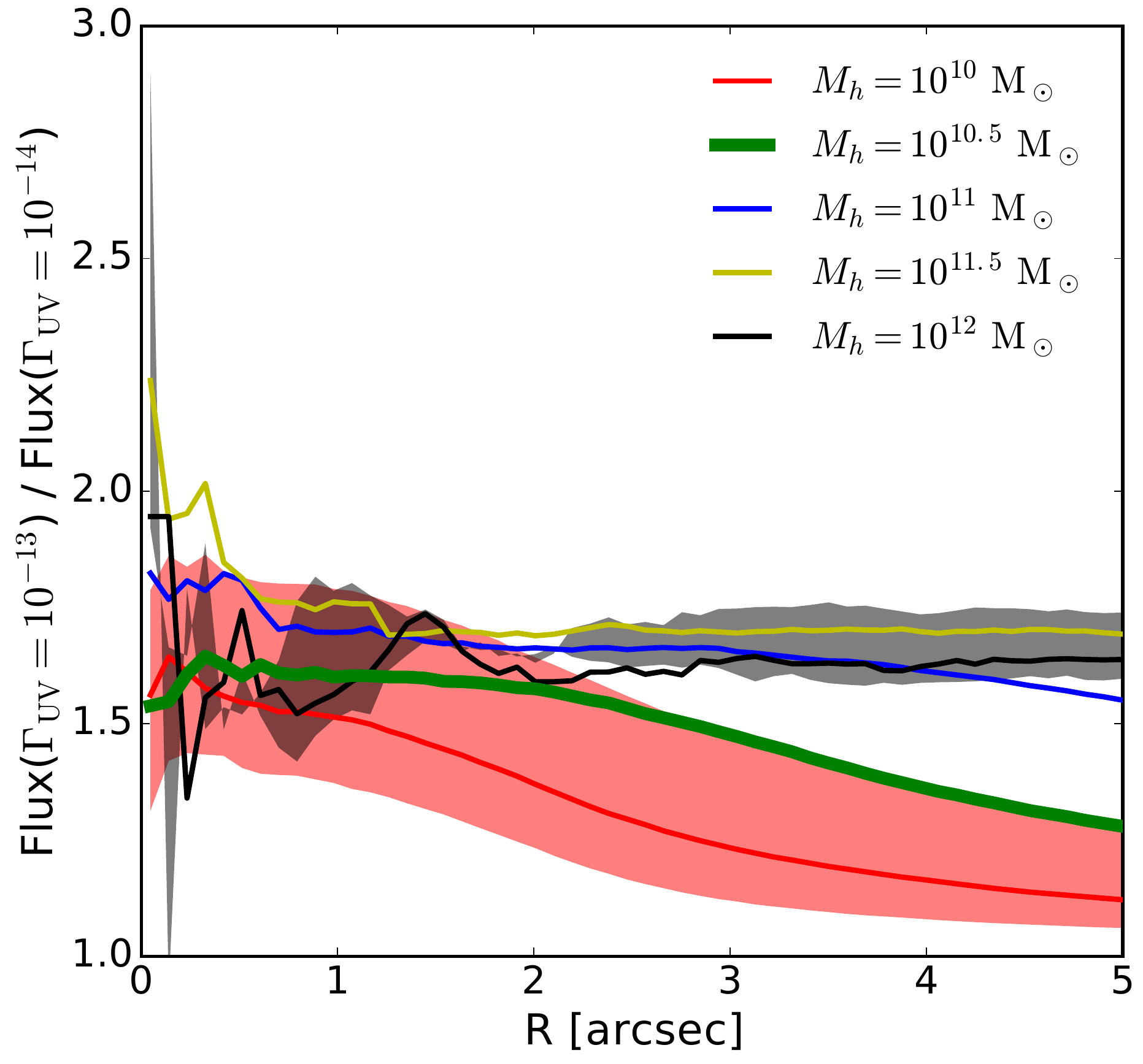}
\caption{
\lya flux ratio as a function of aperture radius.
The fluxes are computed by integrating the surface brightness profiles in 
Figure~\ref{fig:sb_profiles} out to $R$.
For clarity, the scatter (shaded regions) caused by variations of
IGM environments are only shown for the lowest and highest halo mass.
At a fixed projected radius, there is a clear dependence of this ratio on 
the host halo mass, which translates into a luminosity-dependent effect 
on the observability of high-$z$ LAEs. 
}
\label{fig:flux_ratio_profiles}
\end{figure}

\begin{figure*}
\epsscale{1.1}
\plotone{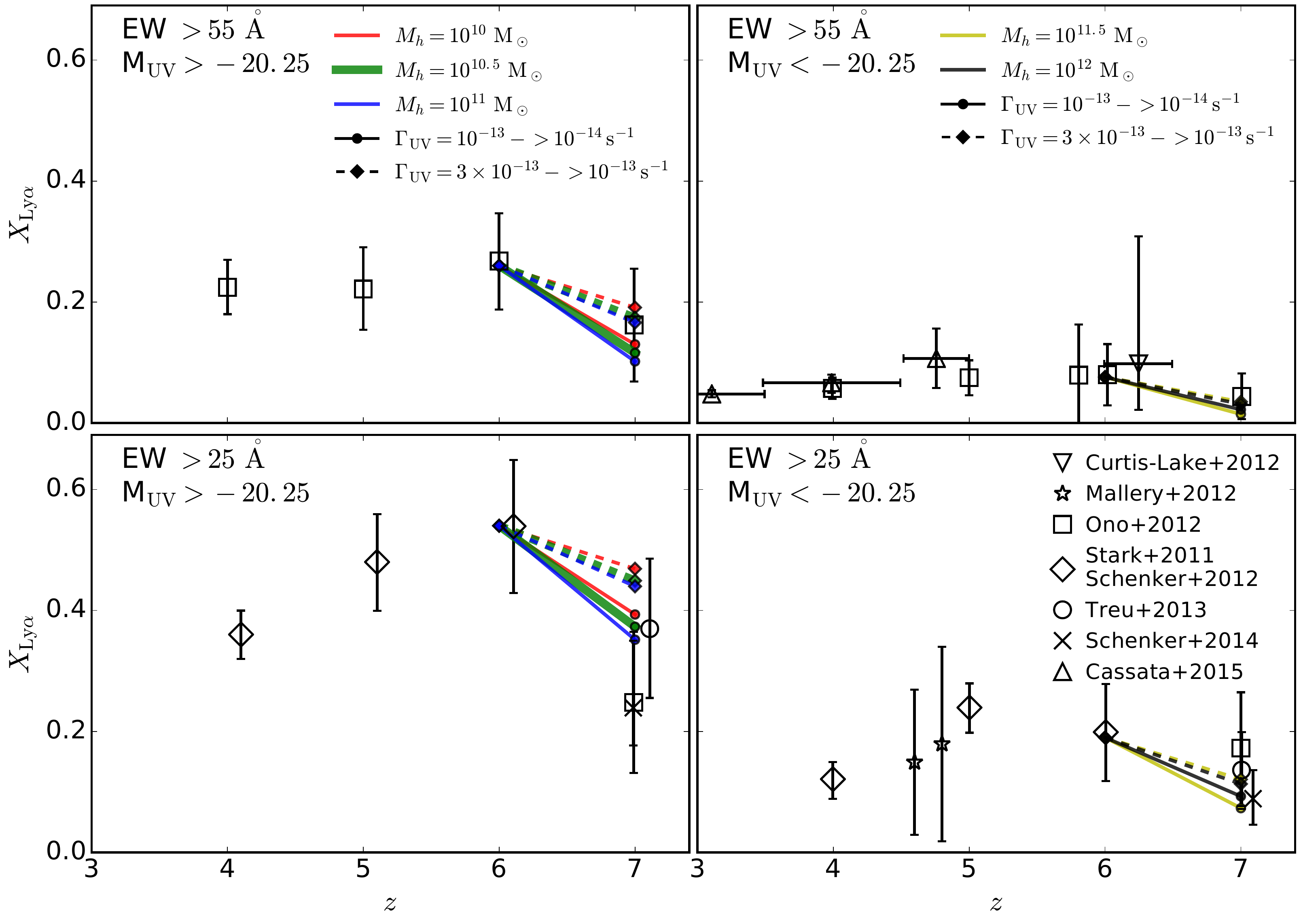}
\caption{
Fraction of galaxies with a \lya equivalent width above the indicated
threshold as a function of redshift. Open black symbols are the
observational data points, and colored filled symbols are predictions
from our LAE model. Observations are taken from 
\citet{Stark.etal11}, \citet{Curtis-Lake12}, \citet{Ono.etal12}, \citet{Mallery12}, 
\citet{Schenker.etal12}, \citet{Treu13}, \citet{Schenker14}, and \citet{Cassata15}. 
\revision[Our model is calibrated to fit the data at $z=6$ with the chosen value 
of the photoionization rate $\Gamma_{\rm UV}$ (so the agreement with data at this redshift is by construction).
We show the prediction of our model for two different redshift evolution of 
$\Gamma_{\rm UV}$ between $z=6$ and $z=7$: from $10^{-13} \, {\rm s}^{-1}$ to $10^{-14} \, {\rm s}^{-1}$ (solid lines) 
and from $3\times 10^{-13} \, {\rm s}^{-1}$ to $10^{-13} \, {\rm s}^{-1}$ (dashed lines).]
The decline in the \lya fraction becomes faster as $\Gamma_{\rm UV}$ is made to also decline faster with
redshift. Results are shown for low (bottom) and high (top) \lya
equivalent width thresholds and for UV-faint (left) and UV-bright (right)
samples. 
}
\label{fig:fLya}
\end{figure*}

  The dependence of the ratio of \lya fluxes for the two cases
($\Gamma_{\rm UV}=10^{-13}\,\mathrm{s}^{-1}$ and $10^{-14}\,\mathrm{s}^{-1}$)
on the aperture radius $R$ is shown in
Figure~\ref{fig:flux_ratio_profiles}. The total flux is computed by
integrating the surface brightness profile out to $R$.
For all halo masses, we find higher flux ratios at smaller aperture radii. 
With an aperture of $1\arcsec$ radius, 
the flux ratios are in the range \revision[$\sim 1.5-1.8$] and show a dependence
on halo mass. The dependence of the flux ratio on halo mass
shows \revision[a weak but interesting trend:] for $M_{h} \la 10^{11}\, \mathrm{M}_\sun$, 
the flux ratio increases with halo mass, but this trend is reversed at higher masses.
This trend may help explain the observed evolution of the \lya fraction
at different galaxy luminosities, which we discuss next.

\subsection{Ly$\alpha$ Fraction Evolution}
\label{subsec:lya_fraction}

  Our model predicts an apparent Ly$\alpha$ flux decrease of LAEs in a
small redshift interval caused by a fast change of the ionizing
background intensity, which results in a rapid shrinkage of
self-shielded regions in the infall zones around the host halos of LAEs.
We now study how this affects the evolution of the Ly$\alpha$ fraction,
$X_{\rm Ly\alpha}$, defined as the fraction of star-forming galaxies
with a \lya equivalent width above a certain threshold. As mentioned in
the introduction, this fraction has been found to gradually increase
up to $z=6$, and then decline suddenly from $z=6$ to $7$
(see the compilation of data points in Figure~\ref{fig:fLya}). 
The Ly$\alpha$ fraction is expressed in terms of the equivalent width
probability distribution $p(W)$ and the threshold $W_t$, as
\begin{equation}
X_{\rm Ly\alpha} = \int_{W_t}^{+\infty}p(W)\, \mathrm{d}W ~.
\label{eq:fLya}
\end{equation}
For a direct comparison with observations, we consider two threshold
values that have commonly been used, $W_t=25$ \AA\ and $W_t=55$ \AA. 
We also adopt the following parametrization for $p(W)$, motivated by
the observed distribution at low redshifts
(e.g., from $z\sim 3$ LBGs; \citealt{Shapley.etal03}):
\begin{equation}
  p(W) = \left\{
     \begin{array}{ll}
       0 & \text{if $W \leq -W_1$},\\
       1/(W_0+W_1) & \text{if $-W_1 < W \leq 0$}, \\
       {\rm exp}(-W/W_0)/(W_0+W_1) & \text{if $W > 0$}~,
     \end{array}
   \right.
\end{equation}
where $W_0$ and $W_1$ are two free parameters (note that $W_1$ is
usually positive since a fraction of the galaxies have negative
equivalent widths).

  We now assume that $\Gamma_{\rm UV}$ drops from the high value of
$10^{-13}\, {\rm s}^{-1}$ at $z=6$, to the low value
$10^{-14}\, {\rm s}^{-1}$ at $z=7$, as an explanation of the fast drop
in $X_{\rm Ly\alpha}$ between these two redshifts.
We determine the values of $W_0$ and $W_1$ at $z=6$ from
the measured values of $X_{\rm Ly\alpha}$ at this redshift, using the
two thresholds $W_t=25$ \AA\ and $W_t=55$ \AA.
Hence, our model matches the data exactly at $z=6$ by construction,
and then predicts the drop at $z=7$ as a function of the value of
$\Gamma_{\rm UV}$.

  Figure~\ref{fig:fLya} shows the change of $X_{\rm Ly\alpha}$
from $z=6$ to $z=7$ in this model, compared to the observations, for the
two equivalent width thresholds. Observational data points are taken 
from \citet{Stark.etal11}, \citet{Curtis-Lake12}, \citet{Ono.etal12}, 
\citet{Mallery12}, \citet{Schenker.etal12}, \citet{Treu13}, 
\citet{Schenker14}, and \citet{Cassata15}, 
which provide constraints on $X_{\rm Ly\alpha}$ for both UV-faint ($\mathrm{M}_\mathrm{UV} >-20.25$) 
and UV-bright ($\mathrm{M}_\mathrm{UV} <-20.25$) galaxy samples. To compare the Ly$\alpha$ fraction 
evolution predicted by our model for these two samples, we therefore need to estimate the 
UV luminosity $L_\mathrm{UV}$ at a given halo mass. For this purpose, we adopt the $L_{\mathrm{UV}}$-SFR relation in \citet{Zheng.etal10} [their equation~(5)]
\begin{equation}
L_\mathrm{UV} = 8\times\,10^{27}\,[\mathrm{SFR}/(\mathrm{M}_\sun \mathrm{yr}^{-1})]\, \mathrm{erg}\, \mathrm{s}^{-1}\, \mathrm{Hz}^{-1},
\label{eq:LUV-SFR}
\end{equation}
where the SFR is related to halo mass $M_{h}$ using equation~(\ref{eq:sfr_Mh}). 
We find that LAEs in halos of mass $M_{h} \le 10^{11}\,\Msun$ 
and $M_{h}>10^{11}\,\Msun$ correspond respectively to the UV-faint and UV-bright sample.
It is worth mentioning that the $L_{\mathrm{UV}}$-$M_h$ relation we adopt here
is only used to give us a rough idea on how to separate our LAE models into UV-faint and UV-bright samples
in order to compare with observations and, for this purpose, is consistent with the relation
inferred from abundance matching at $z\sim 6-7$ \citep{Trac.etal15}.

\revision[With the assumed factor of 10 drop in $\Gamma_{\rm UV}$, we 
find that our model predicts a decline of a factor of 2 or less in $X_{\rm Ly\alpha}$ from $z=6$ to $z=7$, 
in good agreement with observations for both UV-faint and UV-brigth galaxies. 
Given that the redshift evolution of $\Gamma_{\rm UV}$ is still poorly constrained at $z \ga 6$, 
we also show in Figure~\ref{fig:fLya} the prediction of our model using a more moderate 
decline in $\Gamma_{\rm UV}$ from $3\times\,10^{-13}\, {\rm s}^{-1}$ at $z=6$ to 
$10^{-13}\, {\rm s}^{-1}$ at $z=7$ (dashed color lines). In this case, the drop in $X_{\rm Ly\alpha}$ is 
reduced but is still consistent with the observed evolution, especially for halos with $M_{h}<10^{11}\,\Msun$ (UV-faint LAEs) 
which, as we mentioned before, are the ones most affected by a change in $\Gamma_{\rm UV}$. 
By varying the rate of change of $\Gamma_{\rm UV}$ with redshift, we can adjust
our model prediction to the observed evolution of $X_{\rm Ly\alpha}$.]

  We note that although $\Gamma_{\rm UV}$ can continue to increase with
decreasing redshift at $z<6$, the \lya emission equivalent widths may no
longer increase with $\Gamma_{\rm UV}$ once the infall region has been mostly
ionized, if the remaining self-shielded gas is already located at a
radius comparable to the aperture for the observed spectrum.
Furthermore, the gas closer to the center may be ionized by the central
source or by shock heating in the halo, rather than from the external
ionizing background, particularly in massive halos, making the
measured intensity of the \lya emission line insensitive to the value
of $\Gamma_{\rm UV}$ above a value $\sim 10^{-13}\, {\rm s}^{-1}$, which is
required to ionize the infall region. 

\section{Summary and conclusion}
\label{sec:conclusion}

 We construct a simple analytical model to describe the density and velocity 
distribution of the gas around high-redshift LAEs as a function of their 
host halo mass and redshift. The gas distribution is represented by a 
spherically symmetric cloud, consisting of the NFW profile with a core
region, surrounded by an infall region and the IGM in Hubble expansion
farther away from the LAE. Self-shielding on the gas distribution is
computed for two values of the external ionizing background intensity,
which can increase rapidly as reionization proceeds. Based on detailed
\lya radiative transfer calculations, we find that this model is able to
account for the observed decrease in the fraction of \lya
emitting galaxies in the interval from $z=6$ to $z=7$ for both UV-bright
and UV-faint galaxies, if the background intensity drops moderately
\revision[(by a factor $\sim 3-10$)] over this redshift interval.

  The mechanism of this model is that the rapidly growing ionizing
background intensity toward $z\sim 6$ leads to a rapid ionization of the 
infall region surrounding LAE host halos, which greatly reduces the scattering of
\lya photons in this region. As a result, compared to the $z\sim 6$
LAEs, the \lya photons in the red peak that are able to escape and
can produce the observed \lya emission line are much more spatially
extended for $z\sim 7$ LAEs, and the detectable \lya flux within the
small central region corresponding to the commonly used observing
apertures drops by a factor of a few. This provides a natural
explanation for the drop in the fraction of
galaxies with strong \lya emission. 

In this model, a uniform external ionizing background is adopted. During
reionization, however, the UV background is expected to be highly 
inhomogeneous owing to the complex topology of the reionization process and 
the discreteness of ionizing sources 
\citep{Miralda-Escude.etal00,Davies.Furlanetto.16}. 
Inside the regions of the IGM that are highly ionized (i.e., ${\rm H\, II}$ 
regions), the local UV background can reach higher intensities than that
in the neutral IGM far away from ionizing sources. This mainly happens 
at the early stages of reionization, before ${\rm H}\,{\rm II}$ regions have 
overlapped. In the post-overlap phase of reionization,
when all the low-density IGM is ionized, the spatial 
fluctuations in the UV background intensity are reduced, as the mean free 
path of ionizing photons increases rapidly \citep{Gnedin2000}. 
In our model, we set the IGM environment close to LAEs as a neutral 
self-shielded gas, whose spatial extent is determined by the intensity of the 
local UV background, embedded in a large-scale ionized region. This 
corresponds to the late stage of reionization, which proceeds 
\emph{outside-in} after the overlap of ${\rm H\, II}$ regions.
Therefore, we believe that our adoption of a uniform background is a
reasonable approximation around LAEs at $z\sim 6-7$, but of course
improved predictions can be achieved with fully three-dimensional
radiative transfer cosmological simulations. 

  We have not attempted to model the gas distribution inside the halo
virial radius. We have simply introduced a core radius in the neutral
gas distribution, ignoring the effects of complex physical processes
of shock heating or internal ionization and winds, arguing that the
radiative transfer process that matters for the observable \lya emission
line at the redshifts of interest occurs in the infall region and not
within the virialized halo. We have also assumed a turbulent dispersion
and a smooth gas distribution, and have not modeled the effects of
gas inflow and outflow, a possible multiphase distribution including clumps 
\citep[e.g.,][]{Dijkstra12,Duval14}, and an anisotropic gas distribution 
\citep[e.g.,][]{Zheng.Wallace14}. All these factors can affect the \lya 
radiative transfer and can produce anisotropic \lya emission, which may
modify the \lya EW distribution. Our model focusing on the infall and IGM 
regions can be regarded as describing an average effect on the transfer of 
photons escaping the host halo. For future work, a detailed investigation 
with a more realistic gas distribution can come from modeling LAEs in 
high-resolution hydrodynamic galaxy formation simulations.

\revision[As a test for the uncertainty associated with
the gas distribution inside halos, we have performed additional test runs 
by varying the key physical parameters that affect the neutral gas distribution 
within $r<r_h$, namely the size of the constant density core and the 
escape fraction of ionizing photons emitted by the LAE (see Appendix). 
These tests have shown that even when the neutral gas content inside the halo 
is much lower than in the fiducial case, our model still predicts a \lya flux 
decrement from $z\sim 6$ to $7$ at a level similar to that in the
fiducial model. This suggests that our main results on the reduced visibility 
of LAEs towards $z\sim 7$ are robust even if there are uncertainties in our 
modeling of the neutral gas distribution within the halo, as long as a variations 
in the UV background intensity are able to affect the ionization state of the gas in the infall 
region due to self-shielding effects.]

\citet{Dijkstra.etal07} also use the infall model of \citet{Barkana04} to 
study the IGM transmission to \lya emission at $z\ga 6$ by modifying a 
starting \lya line profile based on the \lya scattering optical depth at
each frequency (i.e., the $e^{-\tau}$ model). In our work, after a 
self-consistent self-shielding correction, we track the scatterings of 
\lya photons not only inside the halos but also in the infall regions and 
the IGM. The radiative transfer in the infall and IGM regions leads to 
additional frequency and spatial diffusion of \lya photons that the 
$e^{-\tau}$ model cannot capture \citep[e.g.,][]{Zheng.etal10}. While the
results may be qualitatively similar, our model treats in greater detail
the self-shielding and radiative transfer effects.

In our model, the apparent \lya fraction evolution is caused by the changes 
in neutral gas environment around LAEs induced by the rapid evolution in the
UV background intensity. This rapid evolution of the UV background is expected
as the mean free path of ionizing photons is quickly rising as a consequence 
of the reduced number density of optically thick systems towards the end of 
reionization \citep{Miralda-Escude.etal00}. \citet{Bolton.Haehnelt13} also
propose that the Ly$\alpha$ fraction evolution can be explained by the rapid 
change in the UV background level. Their model differs from ours in that they
attribute the reduction of the LAE visibility to the generally increased
number density of self-shielded regions from $z\sim 6$ to $7$, whereas we
propose that the main effect is due to the change in self-shielding of the
infall regions around the host halos of the LAEs themselves. 
The models may be to some extent complementary and a 
combination of both reasons may provide a more complete picture on the \lya
fraction evolution at $z \ga 6$ \citep[e.g.,][]{Dijkstra14}.

\acknowledgments

We thank Masami Ouchi and Mark Dijkstra for useful discussions.
This work was supported by NSF grant AST-1208891 and NASA grant NNX14AC89G for
RS and ZZ, and by Spanish grant AYA2012-33938 for JM. 
The support and resources from the Center for High Performance Computing at the University of Utah are gratefully acknowledged. 

\bibliographystyle{aasjournal}
\bibliography{paper}

\appendix

\section{Model uncertainties}
\label{sec:appendix}

Our description of the the neutral gas distribution around a typical LAE at $z \ga 6$ (section \ref{sec:model}) 
relies on simplifying assumptions and the physical parameters of our model are poorly constrained at these 
redshifts. Since the \lya radiative transfer depends critically on the neutral gas distribution around the LAE, 
these uncertainties in our modeling can in principle affect the resulting \lya properties we predict. We study this effect 
by varying the value of various physical parameters of our model in order to quantify to what extent the uncertainties 
affect our final conclusion on the evolution of the \lya fraction. 

We specifically investigate the dependence on the core radius $r_{\rm core}$ as well as the escape fraction $f_{\rm esc}$ of ionizing 
photons emitted from the LAE. Given the computational cost associated with the \lya radiative transfer, we perform only two test 
runs by increasing the value of either $r_{\rm core}$ or $f_{\rm esc}$ by a factor of two with respect to their 
fiducial values and only consider a halo of mass $M_{h}=10^{10.5}\,\mathrm{M}_\sun$. 
For the case of a different core radius, we thus adopt the new value of $r_{\rm core} = 0.5 r_{h}$ while we 
increase $f_{\rm esc}$ to $0.2$ when testing the dependence on the escape fraction. 
In addition, we also run an additional test to examine the impact of not including self-consistently the ionizing flux from the central source in 
the self-shielding correction when solving the photoionization equilibrium equation ($\Gamma_{\rm LAE} = 0$ in equation \ref{eq:photoequi}). 
Instead, for this last case, we adopt a larger core radius of $r_{\rm core} = 2/3 r_{h}$ to approximately account for the 
modified neutral gas distribution inside the halo with respect to the fiducial case. 

For each of these three situations, $r_{\rm core} = 0.5 r_{h}$, $f_{\rm esc} = 0.2$ or $\Gamma_{\rm LAE} = 0$, we calculate again the 
corresponding neutral gas distributions, \lya spectra and the evolution of the \lya fraction predicted by our model when 
we vary the UV background intensity from $\Gamma_{\rm UV}=10^{-13}\,\mathrm{s}^{-1}$ at $z=6$ to $\Gamma_{\rm UV}=10^{-14}\,\mathrm{s}^{-1}$ at $z=7$, 
and compare the results with the fiducial case. 

\begin{figure*}
\epsscale{1.2}
\plotone{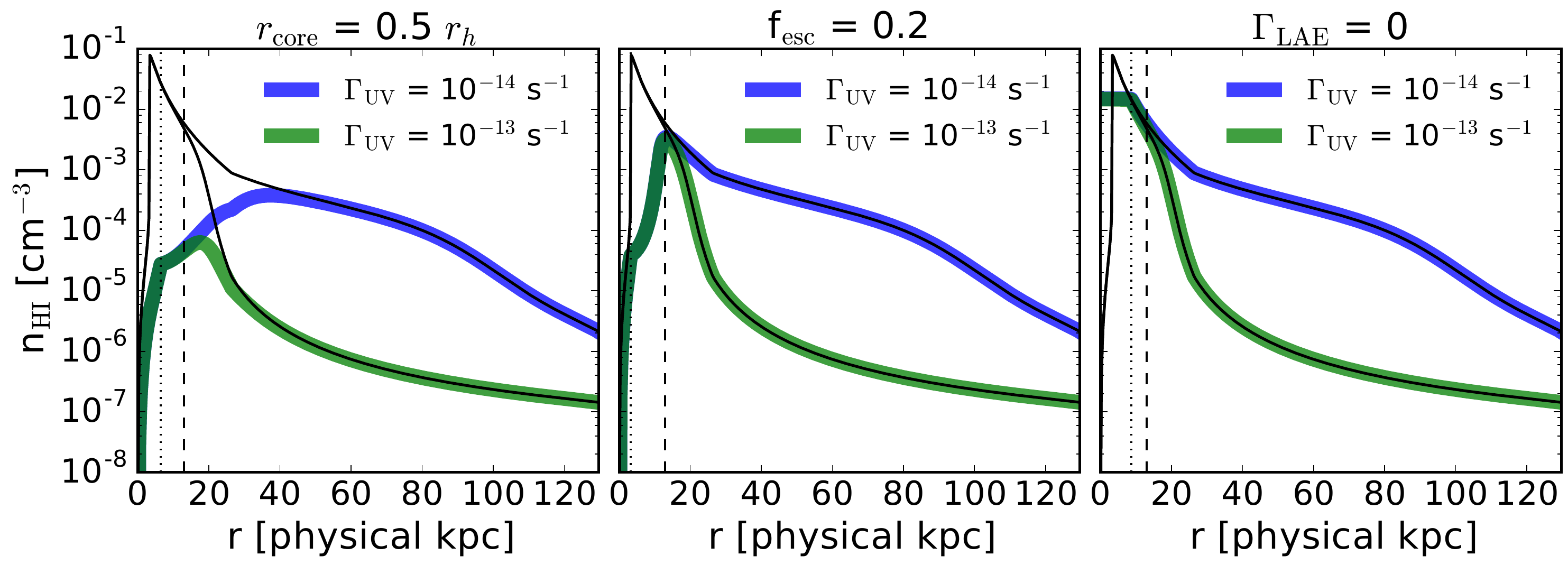}
\plotone{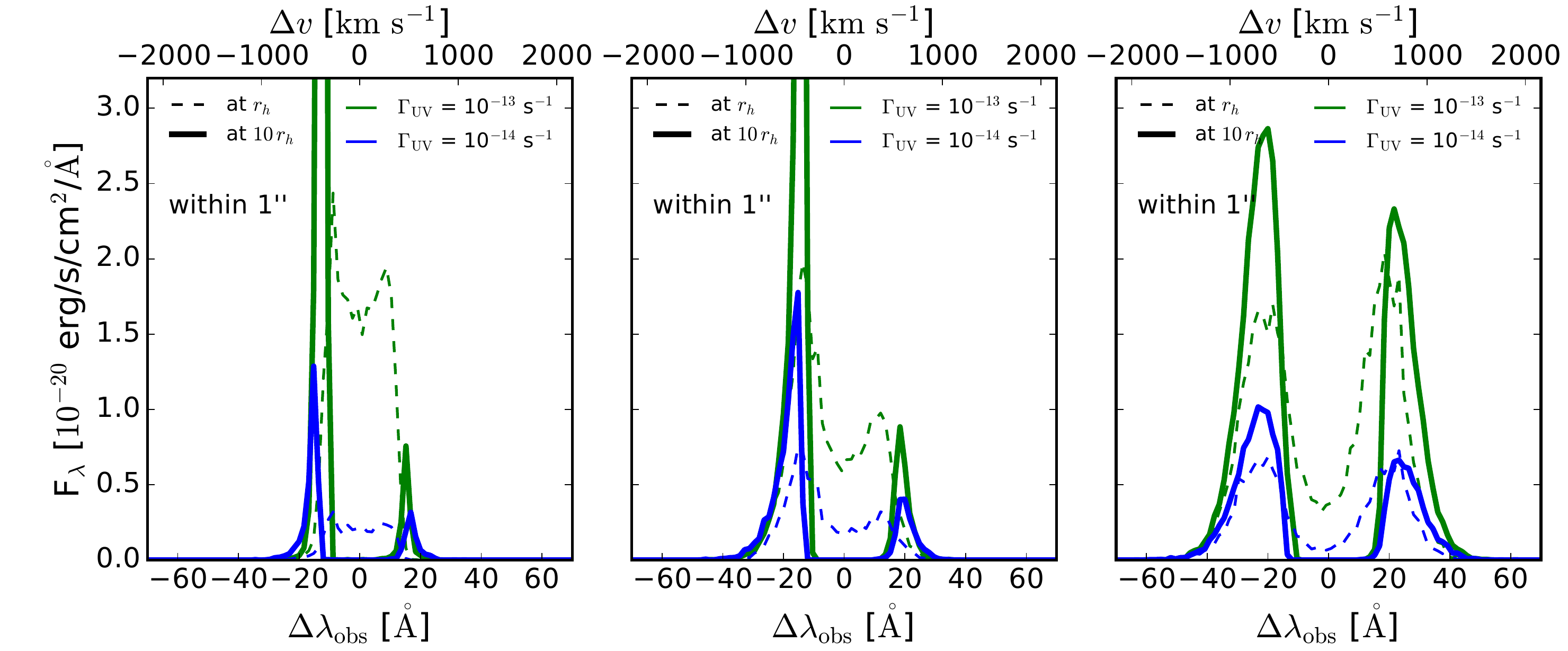}
\caption{
Resulting neutral gas profiles (\emph{top}) and \lya spectra for photons escaping within $1\arcsec$ from the central 
source (\emph{bottom}) after varying some physical parameters of our model for a halo of mass $M_{h}=10^{10.5}\,\mathrm{M}_\sun$. 
The different lines have similar meaning as in Figures \ref{fig:gas_profiles} and \ref{fig:spectra}, except for the solid black 
lines in the top panels which show the neutral gas profile of our fiducial model for comparison (see Fig. \ref{fig:gas_profiles}). 
The first column shows the effect of adopting a larger core radius of $r_{\rm core}=0.5r_{h}$ while the second column 
corresponds to an increase in the escape fraction of ionizing photons from $f_{\rm esc}=0.1$ to $0.2$. 
In the third column, we show the situation where the ionizing flux from the central source is not included 
in the photoionization equilibrium equation ($\Gamma_{\rm LAE} = 0$ in equation \ref{eq:photoequi}). 
Note that for this latter case, we also increase the value of the core radius (see text). 
}
\label{fig:nH_spectra_comparison}
\end{figure*}

\subsection{Effects on the neutral gas distributions}

The top panels of Figure~\ref{fig:nH_spectra_comparison} show the neutral gas profiles obtained for the different 
test runs described above. The black solid lines are the profiles corresponding to the fiducial model presented 
in Figure~\ref{fig:gas_profiles} for comparison. We find that increasing either $r_{\rm core}$ or $f_{\rm esc}$ 
makes it easier for photons emitted from the central LAE source to affect the ionization state of the gas 
out to larger radii as compared to the fiducial model in which these photons are all absorbed at $r<r_{\rm core}$ 
to ionize the gas within the core. Instead, in the the case of a higher $f_{\rm esc}$, the central source is able to maintain 
most of the gas inside the halo ($r<r_{h}$) ionized with a peak neutral gas density $\sim 30$ times 
lower than in the fiducial case. In the case of a larger core radius, photons emitted from the LAE can reach even larger 
radii and ionized the gas out to $\sim 1.5 r_{h}$ simply because the total gas density 
is much lower than in the fiducial model for $r<0.5r_{h}$. In all the test cases however, we find that the ionization state of the 
infall region ($2r_{h}<r<6r_{h}$) is largely unaffected compared to the fiducial situation and is mostly 
governed by the value we assume for the intensity of the external UV background. 

\subsection{Effects on the \lya spectra}

The resulting \lya spectra corresponding to our three test runs are shown in the bottom panels of Figure~\ref{fig:nH_spectra_comparison} 
and the different lines have the same meaning as in Figure~\ref{fig:spectra}. Since the aperture commonly used for LAE detection at 
$z \ga 6$ is $\sim 1\arcsec$ (Section~\ref{subsec:spectra}), we only show the spectra for photons that escape 
within a projected radius of $1\arcsec$ which can be directly compared to the spectra obtained for our 
fiducial model presented in the right panel of Figure~\ref{fig:spectra} (although note the slightly smaller range of the 
y-axis used here). In the cases of a larger $r_{\rm core}$ or $f_{\rm esc}$, the overall main differences seen in the \lya spectra 
compared to the fiducial case are: (1) a smaller offset from the line center of the blue and red peak for both sets of spectra 
emerging from either $r_{h}$ (dashed lines) or $10r_{h}$ (solid lines) and (2) an enhanced asymmetry of the blue peak with respect to the 
red peak in the final spectra emerging at $10r_{h}$. Both of these effects result from the smaller \ion{H}{1} column density inside 
$r<r_{h}$ (top panels of Figure~\ref{fig:nH_spectra_comparison}) which reduces the \lya optical depth and thus the 
number of scattering experienced by \lya photons as they diffuse out of the halo with respect to the fiducial case. 
This is especially visible in the case of $r_{\rm core} =0.5r_{h}$ for which a large number of photons are able 
to escape the halo with frequencies close to the line center and the spectra emerging from $r_{h}$ appear relatively flat. 
As a consequence of the halo being more optically thin to \lya photons, 
the bulk flow of the gas in the infall region (which remains largely unaffected) imprints more effectively the typical asymmetry 
between the blue and the red peak as expected for a source within a collapsing gas cloud \citep[e.g.][]{Zheng.Miralda02b}.

\subsection{Effects on the \lya fraction evolution}

\begin{figure*}
\epsscale{1.2}
\plotone{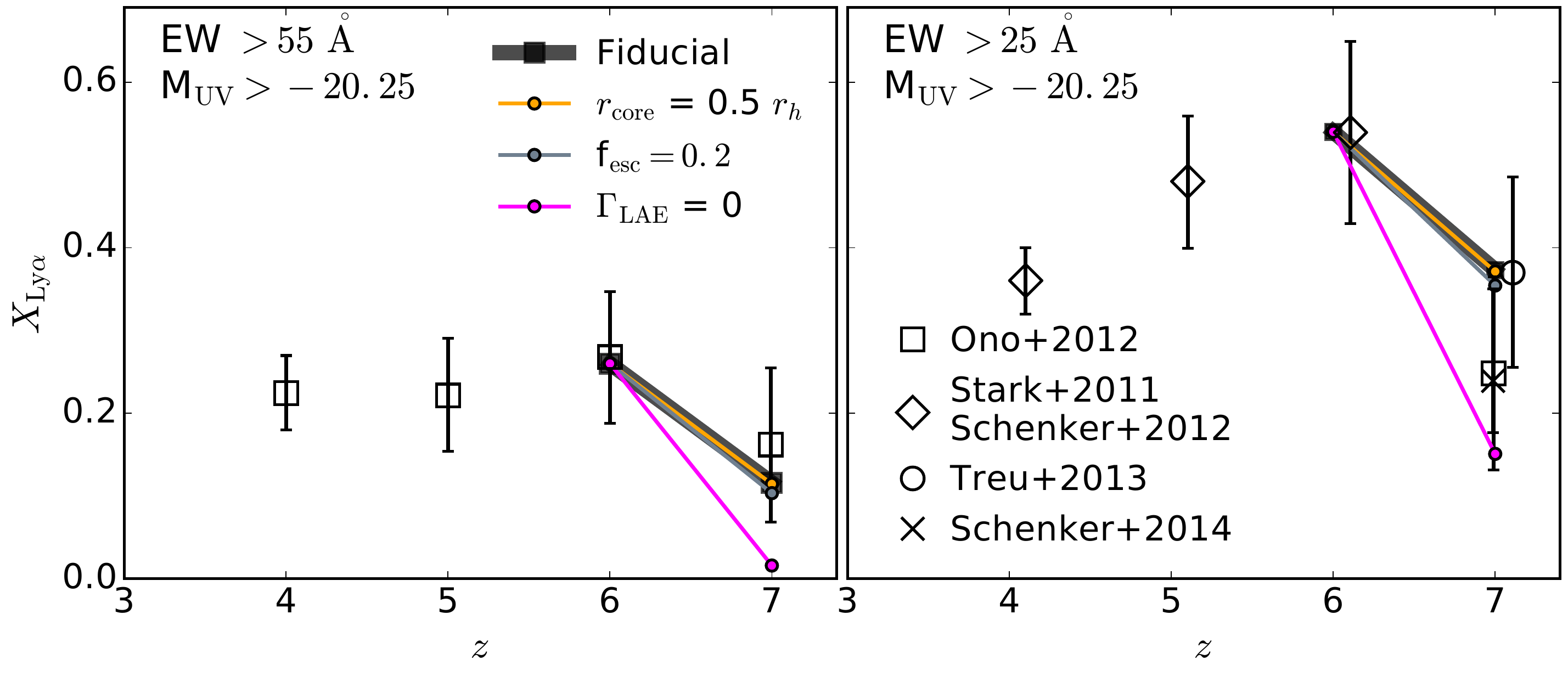}
\caption{
\lya fraction evolution predicted by our model after varying the physical parameters corresponding to the 
different cases shown in Fig. \ref{fig:nH_spectra_comparison} for a halo of mass $M_{h}=10^{10.5}\,\mathrm{M}_\sun$ 
and for the two equivalent width thresholds commonly considered in LAE surveys.
Data points (open black symbols) correspond to the the observed \lya fraction evolution for UV-faint galaxies 
($\mathrm{M}_\mathrm{UV} >-20.25$, as in the left panels of Fig. \ref{fig:fLya}). 
The evolution predicted by our fiducial model is shown as a thick black line with square symbols 
for comparison. Increasing the size of the core or the escape fraction by a factor of 2 has no noticeable effect 
on the \lya fraction evolution indicating that our results seem robust against uncertainties in the modeling.
}
\label{fig:Xlya_comparison}
\end{figure*}

We now examine the redshift evolution of the \lya fraction predicted by our three test runs. The \lya fraction is computed 
using the same method as described in Section~\ref{subsec:lya_fraction} by assuming that the intensity of the ionizing background 
drops from $\Gamma_{\rm UV}=10^{-13}\,\mathrm{s}^{-1}$ at $z=6$ to $\Gamma_{\rm UV}=10^{-14}\,\mathrm{s}^{-1}$ at $z=7$. 
Figure~\ref{fig:Xlya_comparison} shows the resulting \lya fraction evolution for the two equivalent width 
threshold, $W_t=55$ \AA\ (left) and $W_t=25$ \AA (right) for the 3 test cases (thin color lines) 
as well as for the fiducial model (thick black line) for comparison. Observational data points (open black symbols) 
are the same as in the left panels of Figure~\ref{fig:fLya} for UV-faint ($\mathrm{M}_\mathrm{UV} >-20.25$) galaxies 
which corresponds to the halo mass of $M_{h}=10^{10.5}\,\mathrm{M}_\sun$ that we consider here. 
We see that increasing $r_{\rm core}$ or $f_{\rm esc}$ by a factor of 2 still results in the same decline of the 
\lya fraction between $z=6$ and $z=7$ as in the fiducial case, which shows that the drop in \lya flux predicted 
by our model is robust despite the uncertainties in modeling the neutral gas distribution within $r<r_{h}$. 

\end{document}